\documentstyle[aps]{revtex}

\begin{document}
\input psfig

\title{ON THE PATH INTEGRAL LOOP REPRESENTATION OF (2+1) 
LATTICE NON-ABELIAN GAUGE THEORIES}

\author{ J.M. Aroca \\
Departament de Matem\`atiques, Universitat
Polit\`ecnica de Catalunya,\\
Jordi Girona 1 i 3, Mod C-3 Campus Nord,\\
08034 Barcelona, Spain.}

\author{Hugo Fort and Rodolfo Gambini \\
Instituto de F\'{\i}sica, Facultad de Ciencias, \\
Tristan Narvaja 1674, 11200 Montevideo, Uruguay}

\maketitle

\begin{abstract}
A gauge invariant Hamiltonian representation for SU(2) in terms
of a spin network basis is introduced. The vectors of the spin network
basis are independent and the electric part of the  Hamiltonian is
diagonal in this representation. The corresponding path integral for
SU(2) lattice gauge theory is expressed as a sum over colored surfaces,
i.e. only involving the $j_p$ attached to the lattice plaquettes. This
surfaces may be interpreted as the world sheets of the spin networks
In 2+1 dimensions, this can be accomplished by working in a lattice dual
to a tetrahedral lattice constructed on a face centered cubic Bravais
lattice. On such a lattice, the integral of gauge variables over
boundaries or singular lines -- which now always bound three coloured
surfaces -- only contributes when four singular lines intersect at one
vertex and can be explicitly computed producing a 6-j or Racah symbol.
We performed a strong coupling expansion for the free energy. 
The convergence of the series expansions is quite 
different from the series expansions 
which were performed in ordinary cubic lattices. 
In the case of 
ordinary cubic lattices
the strong coupling expansions up to the considered truncation number 
of plaquettes have the great majority of their coefficients
positive, while in our case we have almost equal number of 
contributions with both signs. 
Finally, it is discused the connection in the naive coupling limit
between this action and that of the B-F topological field theory
and also with the pure gravity action.
\end{abstract} 

\newpage

\section{Introduction}

The loop approach to abelian quantum gauge theories was introduced
in the early eighties \cite{gt}. Later it was generalized
to the non-abelian Yang-Mills gauge theory \cite{gt2}. This
Hamiltonian method allows to formulate 
gauge theories in terms of their natural
physical excitations: the loops.
The original aim of this general description
of gauge theories was to avoid gauge redundancy 
working directly in the space of the
gauge invariant excitations. However, soon it was realized that the loop
formalism goes far beyond of a simple gauge invariant description.
The introduction by Ashtekar \cite{a} of a new set of variables
that cast general relativity in the same language as 
gauge theories allowed to apply loop techniques as a 
natural non-perturbative description of Einstein's theory.
In particular, the loop representation appeared as the most
appealing application of the loop techniques to this problem
\cite{rs}.

Recently a Lagrangian approach in terms of loops has been developed for
the U(1) model \cite{abf}, and generalized to include matter fields 
\cite{abfs}.
The resulting action is proportional to the quadratic area of the loop
worldsheet. This allows for Monte Carlo simulations in a more efficient way
than by using the gauge potentials as variables.
While in the abelian case the usual Hamiltonian
in the loop representation can be deduced from the loop action by means of
the transfer matrix analisis, the relation between the two
approaches is more obscure in the non-abelian case. 
In fact, the type of surfaces that appear in the Lagrangian loop
formulation suggest that the pass to the Hamiltonian
can be made simpler if we take a representation
different from the loop representation but that shares
important features like the use of gauge invariant
geometrical objects known as ``spin networks''\cite{RoSm}.  A spin network
basis may be obtained by considering  linear combinations of loops.
While the loop basis is overcomplete and therefore is constrained by a
set of identities known as the Mandelstam identities, the vectors of
spin network basis are independent. Furthermore the electric part of the
Yang-Mills Hamiltonian is diagonal in spin network space.

A path integral formulation of $SU(2)$ gauge theories in terms of the
worldsheets swept out by spin networks has been developed in \cite{rei} and
\cite{afg96}. The worldsheets are branched, coloured surfaces, known in
the mathematical literature as spines \cite{KL}. Such surfaces were
introduced in the seventies \cite{id} in the context of strong coupling
calculations in Yang-Mills theory. The starting point was a character
expantion of the Wilson action. This expantion leads after integration
on the link variables to  a sum of contributions proportional to the
coupling constant raised to a power proportional to the area of closed
coloured surfaces. Each coloured surface corresponds to a diagram having
a certain number of plaquettes.  In the usual hypercubic lattice
formulation \cite{id} the higher order terms in the expansion become
extremely complicated, discourageing attempts to reformulate gauge
theories directly in terms of the coloured surfaces. The main difficulty
lies in the computation of certain group theoretic factors for each
surface. Nevertheless it is possible to reformulate gauge theories in
terms of the surfaces along these lines \cite{rei}.

In the present paper we show that most of the complications of the group
theoretic factors can be avoided by working on a special class of
non-cubic lattices, namely the duals to tetrahedral lattices. This leads
to an easy calculation of the action of spin network worldsheets, and
also simplifies the explicit calculation of strong coupling expansions.

The loop actions  of the abelian gauge theories are written in terms of
the surfaces swept by the time evolution of the loops.   The explicit
form of the loop actions for lattice non abelian gauge theories is known 
to be related with coloured surfaces \cite{rei},\cite{afg96},
known as spines in the mathematical lenguage \cite{KL}. They are related
with the world sheet swept by the evolution in time of the spin 
networks.  However, a complete Lagrangian lattice formulation associated
with the spin network representation is not available up to now.

Coloured surfaces were introduced in the seventies \cite{id} to study the 
strong coupling regime of Yang-Mills gauge theories.  The starting point 
was a character expantion of the Wilson action. This expantion leads after
integration on the link variables to  a sum of contributions proportional 
to the coupling constant raised to a power proportional to the area of 
closed coloured surfaces. Each coloured surface corresponds to a diagram 
having a certain number of plaquettes. In the usual square lattice
formulation \cite{id} the explicit form of the general term of this infinite
sum is not known and therefore the action cannot be rewritten in terms of 
coloured surfaces. The main difficulty relies in the computation of the 
group theoretic factors of each diagram. In this paper we show that the 
introduction of a tetrahedral lattice allows to find a very simple
expression for these factors and therefore to write the path integral 
corresponding to the loop representation of the non abelian gauge theories.

In section II we discuss the spin network Hamiltonian approach to 
Yang-Mills theory. In section III we study the character expansion of 
the Wilson action and we show why the group factors do not allow to get 
a closed form for the action in terms of coloured surfaces. In section 
IV we show that these 
factors may be simply computed in a tetrahedral lattice where the Wilson 
action and the Heat Kernel action take a very simple form. In section V 
we apply this  action to perform a strong coupling expantion for the 
heat kernel version of the theory and we compute the
free energy density $f$ and the average plaquette $P$.
 Finally, in section VI we conclude with 
some comments on further developments.

\section{Spin network representation}

We consider the pure gauge theory with gauge group $G$ 
semisimple and compact ($G=U(1)$ or $SU(n)$).
We start with the Hilbert space
${\cal H}=\otimes_\ell{\cal H}_\ell$ where ${\cal H}_\ell=L_2(G)$
and $\ell$ denote the links of the lattice . On every
link $\ell$ we take the ``position'' basis $|U_\ell>$ labelled by 
the fundamental representation matrices $U_\ell\in G$.
A basis of ${\cal H}$ is given by the vectors 
$|U>=\otimes_\ell|U_\ell>$. Links are oriented and
$U_{\bar{\ell}}=U_\ell^{-1}$.

A gauge transformation is specified by a group element $V_s$ on
every lattice site $s$. The state $|U>$ is transformed into $|U'>$
where $U'_{\ell} = V_{s_1} U_{\ell} V_{s_2}^{-1}$, $s_1$ and $s_2$ being the origin and
end of $l$. The physical states are those that are invariant under gauge
transformations and they define the physical Hilbert space ${\cal H}_{phys}$.

The position and momentum operators $\hat{U}_\ell$, $\hat{E}^a_\ell$
are defined
\begin{equation}
\hat{U}_\ell|U>=U_\ell|U>
%\label{eq:1}
\end{equation}

\begin{equation}
e^{i\theta^a\hat{E}^a_\ell}|U>=|U'> \, ,
%\label{eq:2}
\end{equation}
where $U'_\ell=e^{i\theta^aT^a}U_\ell$, $T^a$ are generators of the group
satisfying
\begin{equation}
{\rm Tr }
 (T^aT^b)=\frac{\delta^{ab}}{2}
%\label{eq:3}
\end{equation}

\begin{equation}
[T^a, T^b]=ic^{abc}T^c.
%\label{eq:4}
\end{equation}

The reference state $|0>=\int dU |U>$ is gauge invariant and
permits to express any state $|\psi>=\psi (\hat{U})|0>$.
To work in the physical space ${\cal H}_{phys}$ we have to
find a basis $|\psi_\alpha >$, that is, a collection of
appropiate gauge invariant functions $\psi_\alpha (U)$.

One choice is to consider polinomials
\begin{equation}
\prod _\ell (U_\ell)_{a_1b_1}\cdots (U_\ell)_{a_rb_r}
(U_\ell^{-1})_{c_1d_1}\cdots (U_\ell^{-1})_{c_sd_s}.
\label{eq:5}
\end{equation}

Imposing gauge invariance leads to contraction of indices
of incoming and outcoming matrices at every site. The result
is the loop representation. That is, states
\begin{equation}
|C_1,C_2,\ldots ,C_M>=W(C_1)W(C_2)\cdots W(C_M)|0>,
\label{eq:6}
\end{equation}
where $C_i$ are closed loops and
\begin{equation}
W(C)={\rm Tr }\prod_{\ell\in C} U_\ell.
\label{eq:7}
\end{equation}

These states generate ${\cal H}_{phys}$ but they are not independent.
They satisfy the so called Mandelstam constraints which for $SU(2)$
read
\begin{equation}
W(C_1)W(C_2)=W(C_1C_2)+W(C_1\overline{C}_2),
\label{eq:8}
\end{equation}
for every two loops $C_1, C_2$ with the same origin.

Instead of (\ref{eq:5}) we can consider the states given by
functions
\begin{equation}
\prod_\ell D^{(\mu_\ell)}_{a_\ell b_\ell}(U_\ell) \, ,
\label{eq:9}
\end{equation}
where $\mu$ runs over the irreducible representations of $G$
and $D^{(\mu)}$ are the matrices of these representations.
(Links in the trivial representation are not mentioned
explicitly.)
Looking to a vertex where enter links in the $\mu_1,\ldots ,\mu_r$
representations and leave links in the $\nu_1,\ldots ,\nu_s$
representations we have functions

\begin{equation}
D^{(\mu_1)}_{a_1b_1}(U_1)\cdots D^{(\mu_r)}_{a_rb_r}(U_r)
D^{(\nu_1)}_{c_1d_1}(V_1)\cdots D^{(\nu_s)}_{c_rd_r}(V_r)
\lambda[^{\mu_1}_{b_1}\cdots ^{\mu_r}_{b_r}|^{\nu_1}_{c_1}\cdots 
^{\nu_s}_{c_s}] \, ,
\label{eq:10}
\end{equation}
where $\lambda[\cdots ]$ are coeficients that ensure gauge invariance.
(We sum over repeated latin indices.)
If we perform a gauge transformation with element $g$ at the present
site we see that $\lambda$ must be an invariant tensor

$$
\lambda[^{\mu_1}_{a_1}\cdots ^{\mu_r}_{a_r}|^{\nu_1}_{d_1}\cdots 
^{\nu_s}_{d_s}]=
$$
\begin{equation}
D^{(\mu_1)}_{a_1b_1}(g^{-1})\cdots D^{(\mu_r)}_{a_rb_r}(g^{-1})
D^{(\nu_1)}_{c_1d_1}(g)\cdots D^{(\nu_s)}_{c_rd_r}(g)
\lambda[^{\mu_1}_{b_1}\cdots ^{\mu_r}_{b_r}|^{\nu_1}_{c_1}\cdots 
^{\nu_s}_{c_s}].
%\label{eq:11}
\end{equation}

The $\lambda $ tensors can be computed from the Clebsh-Gordan
coeficients for the group. The condition for $\lambda$ being
non zero is that the decompositions of the product of the
representations entering and the product of the representations 
leaving have common terms. We note also that the maximum number of links
associated to a given site is small and varies with the dimension
of the space so the order of these tensors are bounded.
The representation defined by these states is called {\em spin network
representation}. The states are represented graphically as
oriented paths with branching points where every single line is
labelled by a representation of the group

\begin{center}
\begin{figure}[t]
\hskip 1cm \psfig{figure=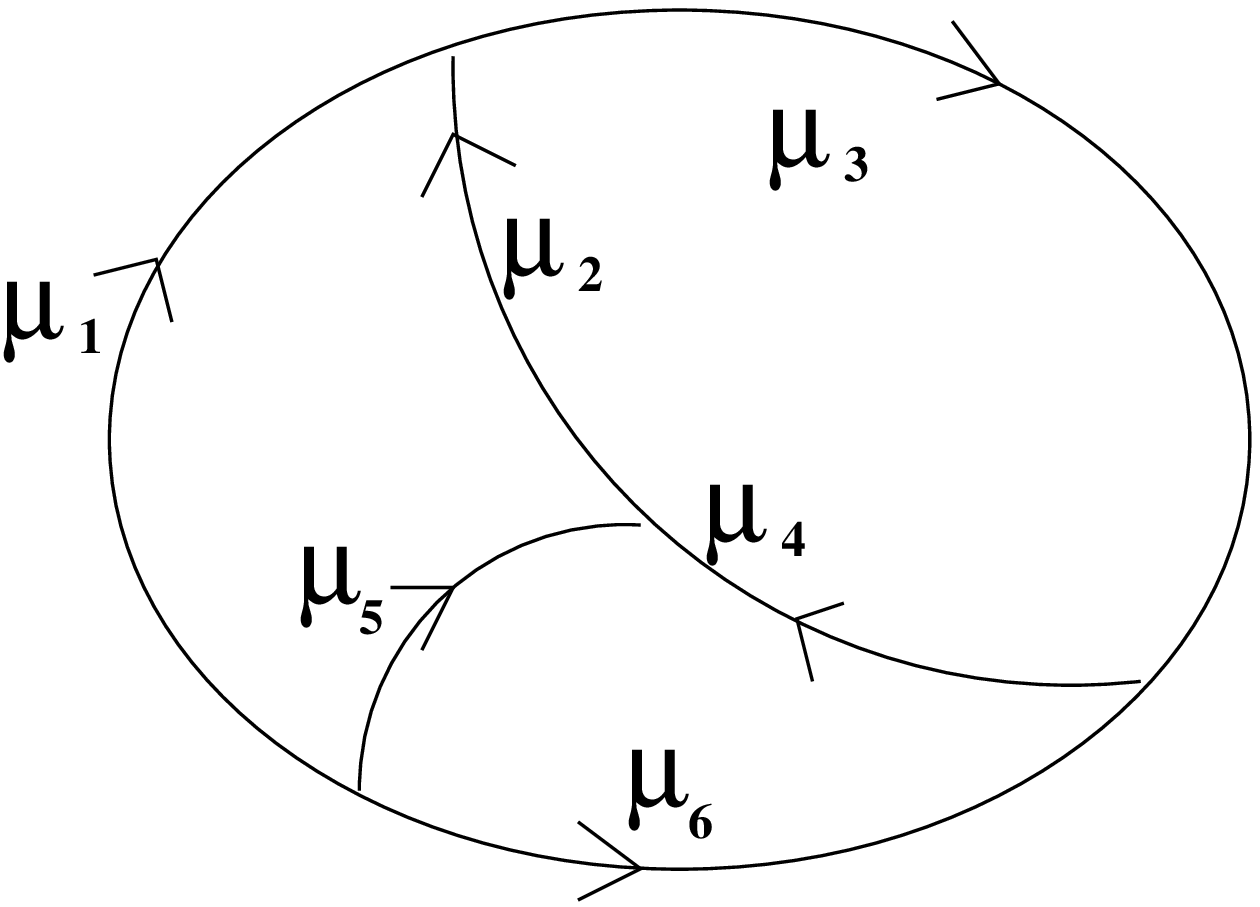,height=5cm}
\caption{}
\label{fig1}
\end{figure}
\end{center}

Supose now that the group is abelian ($G=U(1)$). Then the
irreducible representations are one-dimensional and have the form
$D^{(n)}(U)=U^n$ for $n\in {\bf Z}$. So the (\ref{eq:9}) states 
coincide with the
polinomial states (\ref{eq:5}) and the loop representation and the spin
network representation are the same.

For non abelian $G$ the two representations are different.
It is clear that a site can not have only one link entering
or only one link leaving. The simplest case is one link entering
and another link leaving. Both must be in the same representation
and the invariant tensor is

\begin{equation}
\lambda[^{\mu}_{a}|^{\mu}_{d}]=\delta_{ad}.
%\label{eq:12}
\end{equation}

The efect of this tensor is to multiply the matrices that enter and leave
so single lines in the spin network are labeled by a single representation.
The case of two links entering a site is forbiden in general but
in the case of $SU(2)$ there are invariant tensors like
$\epsilon_{ab}$
(antisymmetric and $\epsilon_{12}=1$). The efect of these tensors is to
reverse one of the lines converging on the site. This is due to
the fact that for $SU(2)$ every representation is equivalent to
its conjugate.

When three lines meet in a site the condition for the existence
of non zero invariant tensors is that the product of the entering
representations contains the trivial representation. In that case
the invariant tensor is given by the CG coeficients

\begin{equation}
D^{(\mu)}_{ab}(g)D^{(\nu)}_{cd}(g)=
\sum_\rho 
C(\mu a,\nu c|\rho d)D^{(\rho)}_{de}(g)C(\rho e|\mu b,\nu d) \, ,
%\label{eq:13}
\end{equation}
which are taken real

\begin{equation} 
C(\mu a,\nu b|\rho c)=C(\rho c|\mu a,\nu b).
%\label{eq:14}
\end{equation}

As we consider only simply reducible groups no other independent
invariant tensor exists. Then the state corresponding to a
trivalent vertex is well encoded in the drawing of the spin network.
This is not in general the case for sites where meet four or more lines.
The dimension of the space of invariant tensors is the number of times
that the product of the entering representations contains the 
trivial representation. Besides the spin network it must be provided
information about which tensors we select on the vertices.
This degeneration is not fundamental since these
vertices can be thought as the limiting case when the line connecting
two trivalent vertices gets zero lenght.

\begin{center}
\begin{figure}[t]
\hskip 1cm \psfig{figure=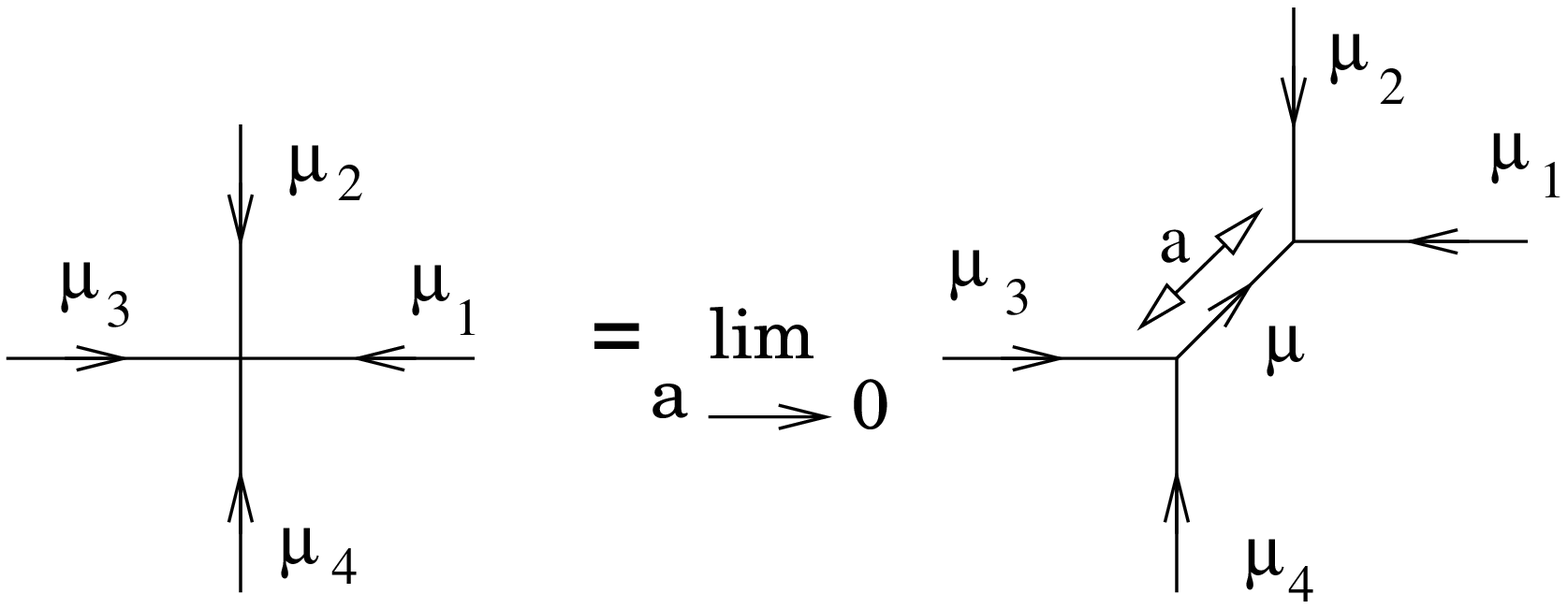,height=4cm}
\caption{}
\label{fig2}
\end{figure}
\end{center}

The Hamiltonian operator is

\begin{equation}
H=\frac{g^2}{2}\sum_\ell \hat{E}_\ell^2-\frac{1}{2g^2}
\sum_p (W(p)+W(\bar{p})) \, ,
%\label{eq:15}
\end{equation}
where $g$ is the coupling constant, $\ell$ and $p$ are the links
and plaquettes of the lattice respectively, 
$\hat{E}_\ell^2=\hat{E}_\ell^a\hat{E}_\ell^a$ and $W(p)$ is given by (\ref{eq:7}).
The fundamental commutation relations are ($T^{(\mu)}_a$ are the generators
of the $\mu$ irred. rep.):

\begin{equation}
[\hat{E}_\ell^a, D^{(\mu)}(\hat{U}_\ell')]=
-T^{(\mu)}_aD^{(\mu)}(\hat{U}_\ell)\delta_{l l'} +
D^{(\mu)}(\hat{U}_\ell')T^{(\mu)}_a\delta_{\bar{l} l'}
%\label{eq:16}
\end{equation}

\begin{equation}
[\hat{E}_\ell^a, \hat{E}_\ell^b]=ic^{abc}\hat{E}_\ell^c.
%\label{eq:17}
\end{equation}

A spin network state is denoted ${\cal N}=P_1P_2\cdots P_M$
where $P_i$ are the single lines. $\mu_i$ is the representation
carried by line $P_i$. We get for the action of the electric part
of the Hamiltonian:

\begin{equation}
\sum_\ell \hat{E}_\ell^2 |{\cal N}>=L({\cal N})|{\cal N}> \, ,
\label{eq:18}
\end{equation}

\begin{equation}
L({\cal N})=\sum_i c^{(\mu_i)}L(P_i) \, ,
%\label{eq:19}
\end{equation}
where $L(P_i)$ is the number of links in the line $P_i$
and $c^{(\mu)}$ is the quadratic Casimir number of the representation
given by $T^{(\mu)}_aT^{(\mu)}_a=c^{(\mu)}Id$.
For example, for $SU(2)$, 
$c^{(j)}=j(j+1)$ .

Then the spin network states, in contrast to the loop states,
are always eigenstates of the electric term.

The magnetic part produces deformations of the spin networks both
in the geometrical shape and the representations of the lines.
This is represented graphically:

\begin{center}
\begin{figure}[t]
\hskip 1cm \psfig{figure=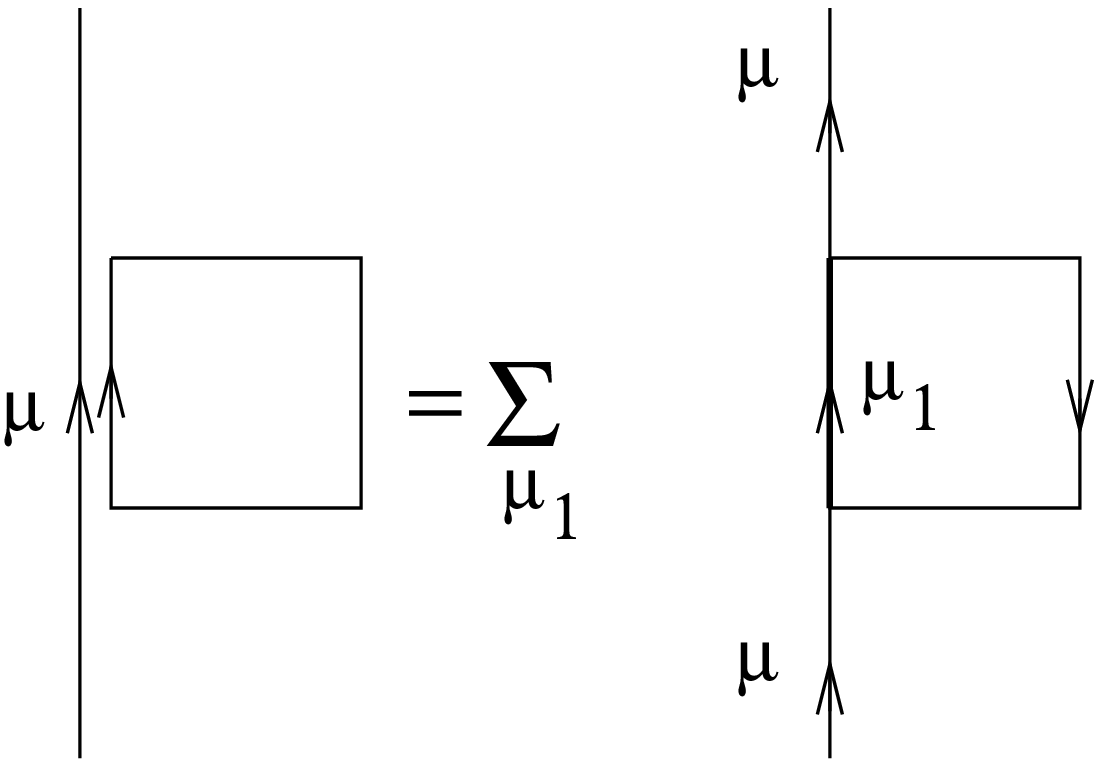,height=4cm}
\caption{}
\label{fig3}
\end{figure}
\end{center}

The spin network states $|\cal{N}>$ are independent and orthonormal.
A proof using an inner 
product structure on the state space is given in \cite{RoSm}. Loop states
can be expanded in spin network terms in a unique way so the Mandelstam
constrains disappear when we pass to the spin network representation.

To further clarify the relation between loop and spin network
states we recall the action of the electric part of the Hamiltonian
over a loop state (\ref{eq:6})

$$
\sum_\ell \hat{E}_\ell^2 |C_1,C_2,\ldots ,C_M>=
$$
$$
(nL(C_1,C_2,\ldots ,C_M)-\frac{1}{n}\Lambda(C_1,C_2,\ldots ,C_M))
|C_1,C_2,\ldots ,C_M>
$$
$$
+\sum_i\sum_{l,l'}\bar{\delta_{l,l'}}
|C_1,\ldots,C_{i xy},C_{i yx},\ldots,C_M>
$$
\begin{equation}
+2\sum_{i<j}\sum_{l\in C_i}\sum_{l'\in C_j}\bar{\delta_{l,l'}}
|C_1,\ldots ,(C_{i xx}C_{j yy}),\ldots,C_M> \, ,
\label{eq:21}
\end{equation}
where $L(C_1,C_2,\ldots ,C_M)$ is the number of links in the set of
loops taking acount of the multiplicity (single lenght),
 and $\Lambda(C_1,C_2,\ldots ,C_M)$ is the sum of the squares of
these multiplicities (quadratic lenght). $\bar{\delta}_{l,l'}$ is 1
if $l=l'$ and $-1$ if $\bar{l}=l'$ and $x,y$ are the edges of $l$.

In equation (\ref{eq:21}) we see that in general the loop states are not
eigenstates of the electric operator. The last two sums are
geometric interaction terms: fusions where a loop splits into
two components and fisions where two loops join in a common link.

\section{The Wilson action in a cubic lattice}

     The path integral for the Wilson
action for a general non-Abelian compact gauge group G is given by
\begin{equation}
Z_W = \int [dU_{\ell} ]\exp [{\beta\sum_p \mbox{Re} 
( {\rm Tr } U_p ) \,]}, 
\label{eq:ZWNA}
\end{equation}
where the $U_\ell \in G$ and $U_p = \prod_{\ell \in p} U_\ell$.

The analogous of the Fourier expansion
for the non-Abelian case is the {\em character} expansion.
The characters $\chi_r(U)$ of the
irreducible (unitary) representation $r$ of dimension
$d_r$, defined as the traces of these
representations, are an orthonormal basis for the
{\em class} functions of the group i.e. \cite{id}
\begin{eqnarray}
\int d U \chi_r(U) \chi_s^*(U)=\delta_{rs} 
\label{eq:ort1} \\
\sum_r d_r \chi_r(U{V^{-1}})=\delta(U,V).
\label{eq:ort2}
\end{eqnarray}
In particular, as a useful consequence we have 
\begin{equation}
d_r \int d U \chi_s(U) \chi_r(U{V^{-1}})=\delta_{rs}\chi_r(V).
\label{eq:ort3}
\end{equation}

By means of the character expansion we can express
\begin{equation}
\exp  \{ \beta\sum_p \mbox{Re} [ {\rm Tr } U_p \,] \}
= \prod_p \sum_r c_r \chi_r (U_p),
\label{charexp}
\end{equation}
with
\begin{equation}
c_r=\int dU \chi_r^*(U)\exp{(\beta {\rm Tr } U\, )}.
\label{eq:coef}
\end{equation}

In the case of $G = SU(2)$ a
direct application of (\ref{eq:coef}) yields the $c_j$
in terms of modified Bessel functions, and therefore
we can express (\ref{eq:ZWNA}) as 
\begin{eqnarray}
Z_W = \int [dU_{\ell} ] \prod_p
[ \sum_{j_p} 2(2j_p+1)\frac{I_{2j_p+1}(\beta)}{\beta}
\chi_{j_p}(U_p)]
\nonumber \\
=\sum_{ \{ j_p \} } \prod_p [ 2(2j_p+1)
\frac{I_{2j_p+1}(\beta)}{\beta} ] \int [dU_\ell] \prod_p
\chi_{j_p}(U_p).
\label{eq:ZWNA2}
\end{eqnarray}

It will be convenient to introduce the following quantities in order to
rewrite the path integral $Z_W$
\begin{equation}
c_0=\frac{2I_1(\beta)}{\beta},
\end{equation}
\begin{equation}
c_0\beta_j=\frac{2I_{2j+1}(\beta)}{\beta}.
\end{equation}
The path integral takes the form,

\begin{equation}
Z_W={c_0}^{N} \int{\prod_p[1+\sum_{j_p \neq 0}(2j_p+1)
 \beta_{j_p}\chi_{j_p}(U_p)]}\prod _\ell dU_\ell
\end{equation}
where $N$ is the number of plaquettes.

Let us now show how the sum over coloured surfaces arise in $Z_W$.
A given subset of plaquettes carring $j_p \neq 0$
is homeomorphic to a simple
surface if any link bounds at most two plaquettes
of this subset. The links bounding exactly one plaquette
make up the free boundary of this surface.
Any configuration can be
decomposed as a set of maximal simple surfaces 
by cutting it along
the links bounding more than two plaquettes.
In principle, there are 
two possibilities for the boundary curves:
either a free boundary, bounding only one simple surface or
a singular branch line along which more than two simple
surfaces meet. In fact, relation (\ref{eq:ort1}) forbids the 
existence of free boundaries for non trivial
configurations contributing to the path integral.

The integration over the internal links of the simple surfaces
is performed using (\ref{eq:ort3}). Note that the plaquettes
of a simple surface component should carry the same  
group representation. After integrating over all the inner links
of the simple components one gets \cite{dz83} an expression
involving only the links of the boundaries,

\begin{equation}
Z_W=c_0^{N}\sum_{\mbox{\small{spines}}}{\int{(\prod_{l\in\partial A}dU_\ell}}) \prod_i 
{\beta_{j_i}}^{A_i}(2j_i+1)^{\epsilon_i}
\prod_{\mbox{\small{boundaries}}}{\chi_{j_i}[\partial A_i]}
\, ,
\label{bound}
\end{equation}
where $\epsilon_i$ is the Euler's topological invariant of the surface $i$
with area $A_i$. Euler's characteristic is explicitly given by:

\begin{equation}
\epsilon= n_2-n_1+n_0= 2- 2g -b
\end{equation}

where $n_2$ is the number of plaquettes, $n_1$ the number of distict links
bordering these plaquettes and $n_0$ the number of end points of these
links, $g$ is the genus of the surface and $b$ the number of boundaries.

 An important property of the character expansion (\ref{eq:ZWNA2}),
relevant for strong coupling expansions is that only a finite number of
terms in \ref{charexp} contribute to a given order in $\beta$. In fact,
\begin{equation}
c_r=O[\beta^\nu_r]
\end{equation}
where $\nu_r$ is the smallest integer such that $\chi^\nu(U)$has a non
vanishing component along $\chi_r(U)$.In the SU(2) case $\nu_j=2j$.

In \cite{rei} and \cite{afg96} a path integral over coloured surfaces is
obtained along the lines described here. However, on a cubic lattice the
group factors of the surfaces are difficult to compute because with up
to four surfaces meeting at singular lines, and up to six singular lines
meeting at points, the integral in (27) can be very complicated,
involving recoupling coefficients of up to 12 $j$s. That is way these
group factors has been only perturbatively computed for the diagrams
that appear in the strong coupling expantion.

Working with a cubic lattice is equivalent to work with
spin networks involving four valent vertices in the Hamiltonian approach
discussed in section II. In this case, it is well known that only three
valent vertices have an unambigous   correspondence with the information
encoded in the drawing of the spin network. Higher order vertices
require additional information about the invariant tensor used to couple
the irreducible representations. At the action level this means that
additional group factors associated with different ways of coupling the
coloured surfaces at singular lines appear. In \cite{rei} this problem
is dealt with by assigning colours to the singular lines which are
summed over in the path integral. However, this complication may be
avoided by using a special class of lattices. In the dual to a
tetrahedral lattice only three plaquettes meet at each link, so singular
lines involve at most three coloured surfaces.

\section{Loop actions in tetrahedral lattices}

        In order to introduce the 
 tetrahedral lattice above mentioned, some concepts of solid state physics are
 very useful \cite{AsMe}. The {\em Bravais} lattice is one of such concepts;
 it specifies the periodic array in which the repeated units of a 
 crystal are arranged. That is, the Bravais lattice summarizes the
 geommetry of the underlying periodic structure, regardeless 
 of what the actual units
 be (single atoms, molecules, groups of atoms, etc.). 
 A (three-dimensional) Bravais lattice is specifyed by
 three vectors {\bf a}$_1$, {\bf a}$_2$ and {\bf a}$_3$ called
 {\em primitive vectors}. The primitive vectors generate all the 
 traslations such that the lattice appears {\em exactly} the same. 
 The primitive unit cell generated by the primitive vectors often does
 not have the full symmetry of the Bravais lattice. However, one can
 always consider a nonprimitive unit cell, known as {\em conventional}
 unit cell, which is generally chosen to be bigger than the primitive
 cell and such that to have the full symmetry of their Bravais lattice.
 
 Le us consider a face centered cubic Bravais lattice -- i.e.
 the lattice obtained when one adds to the simple cubic lattice
 an aditional point in the center of each square face --
 with primitive vectors 
 ${\bf a_1}=a({\bf i+j}), {\bf a_2}=a({\bf j+k}), {\bf a_3}=a({\bf k+i})$
 The conventional unit cell of this lattice is a cube of side $2a$ with a four
 point basis located at $(0,0,0),(a,0,0),(0,a,0)(0,0,a)$ Translations
 along the primitive vectors generate 27 points associated with 8 cubes
 of side $a$ in the conventional cell.The Bravais conventional unit cell
 with the four basis points and the eight cubes is depicted in FIG. 4.

\begin{center}
\begin{figure}[t]
\hskip 1cm \psfig{figure=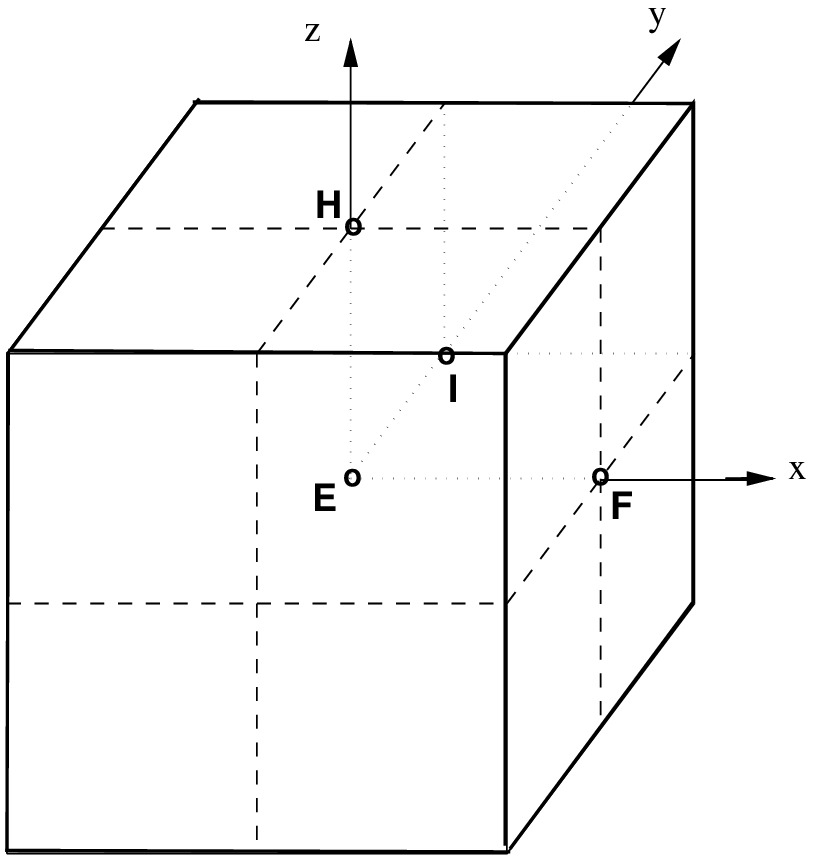,height=4.5cm}
\caption{}
\label{fig4}
\end{figure}
\end{center}

Each cube of side $a$ may be decomposed in the five tetrahedra
$ABDE,\;CBDG,\;EBGE,\;HDGE,$ and $EBGD$ as shown in FIG. 5. The links of
the lattice are the edges of these tetrahedra. The first four
tetrahedra has volume $a^3/6$ while the last one has volume $2a^3/6$.
 If the vertex $A$ of the cube depicted in FIG. 5 has coordinates(0,-a,0)
 the other cubes are obtained by symmetizing with respect to the planes
 $Exy,\;Eyz,\;Ezx$  and traslating along the primitive vectors.
        Coloured surfaces will be associated to the plaquettes of the
 dual lattice.The vertices of this lattice are the centers of the
 tetrahedra:
\begin{eqnarray}
X_A({\bf \epsilon}) = \frac{a}{4}(\epsilon_1,3\epsilon_2,\epsilon_3)\\
X_C({\bf \epsilon}) = \frac{a}{4}(3\epsilon_1,3\epsilon_2,3\epsilon_3)\\
X_F({\bf \epsilon}) = \frac{a}{4}(3\epsilon_1,\epsilon_2,\epsilon_3)\\
X_H({\bf \epsilon}) = \frac{a}{4}(\epsilon_1,\epsilon_2,3\epsilon_3)\\
X_O({\bf \epsilon}) = \frac{a}{2}(\epsilon_1,\epsilon_2,\epsilon_3)\\
\end{eqnarray}
where $\epsilon=\pm1$

\begin{center}
\begin{figure}[t]
\hskip 1cm \psfig{figure=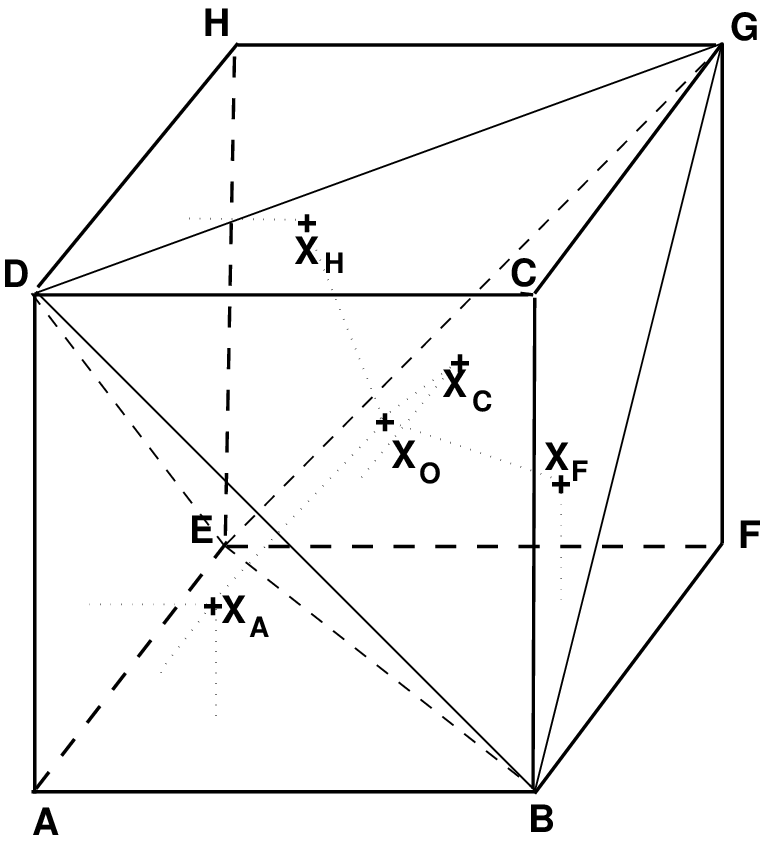,height=5cm}
\caption{}
\label{fig5}
\end{figure}
\end{center}

Each cell contains one polyhedron with 12 hexagonal faces and six
squared faces. In FIG. 6 we show the points and links of one polyhedron
and one cube. Translations along the primitive vectors fill all the
lattice. Each of the squares is a face of one cube of side $a/2$. We
shall attach a $SU(2)$ group element $U_\ell$ in the fundamental
representation to each link of this lattice.

\begin{center}
\begin{figure}[t]
\hskip 1cm \psfig{figure=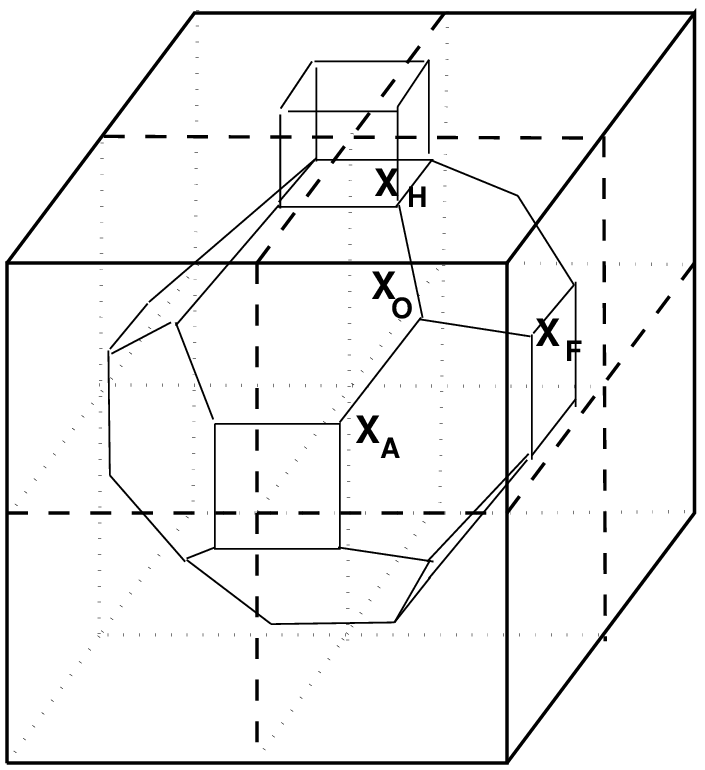,height=6cm}
\caption{}
\label{fig6}
\end{figure}
\end{center}

Now let us consider the Wilson action defined in terms of the plaquettes
of this lattice.
\begin{equation}
S=\beta[\sum_{p_s} {\rm Re}({\rm Tr } U_{p_s})+\sum_{p_h}{\rm Re}({\rm Tr } U_{p_h})]
\end{equation}
where $U_{p_s}$ and $U_{p_h}$ respectively are the holonomies for the
squared and hexagonal plaquettes.

One can show that this action has the correct continum limit for $a$
going to zero, leading to the classical Yang-Mills action.

        One can repeat the same steps leading two (\ref{bound}), but now
the singular lines always bound three coloured surfaces. In this case
the integral along the boundaries in (\ref{bound}) only contributes when
four singular lines intersect at one point and may be explicitly
computed. Let us call $S_k$ the intersecting point and
$\gamma_1,...\gamma_4$ the singular lines intersecting at $S_k$. Then we
have six coloured surfaces with colours $j_{12}, j_{13}, j_{14}, j_{23},
j_{24}$ and $j_{34}$ bounded by these lines. That means that in the
original tetrahedral lattice we shall have a tetrahedrom with one of the
values of $j$ on each edge. The exact path integral may now be written
in terms of a sum over coloured surfaces.
\begin{equation}
Z_W=c_0^{N}\sum_{spines}\prod_i {\beta_{j_i}}^{A_i}(2j_i+1)^{\epsilon_i}
\prod_{S_k}{(-1)^{\sum_i  j_{ii+1}^k}}\left\{ \begin{array}{lll}
                  j_{12}^k,  & j_{13}^k,  & j_{14}^k \\
                  j_{34}^k,  & j_{24}^k,  & j_{23}^k
     \end{array}
     \right\}
\label{spine}
\end{equation}

where the six $j$ symbols are the Racah coefficients and the exponent
of $-1$ denotes the cyclic sum $j_{12}+j_{23}+j_{34}+j_{41}$.

        A simpler but equivalent action  may be obtained
starting from the Heat Kernel path integral

\begin{equation} Z_{HK} =
\int [dU_{\ell} ] \prod_p [ \sum_{j_p} (2j_p+1)\exp
\left\{-\frac{j_p(j_p+1)}{2\beta} \right\} \chi_{j_p}(U_p)] \, .
\end{equation}

By following the same steps one gets the same expression
for the path integral in terms of coloured surfaces (\ref{spine}) with
$c_0=1$ and $\beta_j=\exp\{ -[j(j+1)/2\beta]A_j\}$.

\section{Strong coupling expansions}

 Our aim in this section is to show that the introduction of
the previous Bravais lattice not only allows to perform calculations
but also simplify them. Therefore, we will show here how to perform
a strong coupling expansion.  
In order to do so we will use, just
for simplicity, the
expression for the path
integral in terms of coloured surfaces 
which corresponds to the Heat Kernel action:
\begin{equation}
Z=\sum_{spines}\prod_i \exp\left\{-\frac{j_i(j_i+1)}{2\beta}A_i
\right\}
%^{A_i}
(2j_i+1)^{\epsilon_i}
\prod_{S_k}{(-1)^{\sum_i  j_{ii+1}^k}}\left\{ \begin{array}{lll}
                  j_{12}^k,  & j_{13}^k,  & j_{14}^k \\
                  j_{34}^k,  & j_{24}^k,  & j_{23}^k
     \end{array}
     \right\}.
\label{eq:spine2}
\end{equation}
We will follow an analogous treatment to that of
Drouffe and Zuber \cite{dz83}. We will expand in powers of the 
$t$ parameter given by $t\equiv e^{-\frac{1}{4\beta}}$.

\vspace{4mm}

{\bf Free energy density f}

\vspace{2mm}

The free energy density $f=F/N$, 
where $F$ is the free energy and $N$ the number of plaquettes,
is obtained by summing  
the terms linear in N in the expansion of the path integral 
(\ref{eq:spine2}) in powers of $t$.
The power of $t$ of each diagram (volumes in 
three space-time dimensions) is equal to 

$$\sum_r 2j_r(j_r+1) \times n_r$$

where $j_r$ denotes the representations of the 
group SU(2) or ``colours''
and $n_r$ denotes the number of plaquettes 
(square plaquettes + hexagonal plaquettes) 
of the the diagram. 
For instance, the first power of the expansion corresponds to the 
smallest volume, i.e. the cube, with all their plaquettes 
with $j=\frac{1}{2}$ and it gives a power of 2$\times \frac{3}{4}
\times 6 = 9$; the next power is produced by two disconnected
cubes (recall that in our lattice cubes make contact only
with polyedra) with  $j=\frac{1}{2}$ which gives a power of 18 and
so forth. The contribution of each diagram to $f$ can be
written as the product of two numbers: the {\em reduced 
configuration number} (r.c.n.) times a group theoretical factor
\cite{dz83}. To compute the r.c.n. one has to count
the number of inequivalent positions of a given diagram 
on the lattice -- its {\em configuration number} --
and then to extract the term linear in $N$ which is the r.c.n.
The group theoretical factors stem from the integrations over
the link variables $U_\ell$ and their general form is

$$\sum_{r \ne 0} d_r^{n_2-n_1+n_0} \prod_{\mbox{\small{boundaries}}} 
\chi_r(U_{\mbox{\small{boundary}}}),$$   

where $d_r$ is the dimension of the representation $r$,
$n_2$ is the number of plaquettes with $j=j_r$,
$n_1$ the number of distinct links bordering these plaquettes
and $n_0$ the number of endpoints of these links.
For example, diagrams with the topology of a sphere
give contributions $\sum_{r \ne 0} d_r^2$. The main advantage of
the introduced lattice is that the group theoretical factors 
for more complicated 
diagrams can always be explicitely expressed in terms of
the $d_r$ and 6-$j$ Racah symbols.
The Racah symbols arise each time four singular lines 
meet at one vertex and they appear in the diagrams by pairs. 
The first of this pairs come out in the diagram of two 
polyhedron sharing an hexagonal face and a
cube sharing two of its contiguous faces, one 
with each polyhedron (FIG. 7). 
\vspace{3mm}

\begin{center}
\begin{figure}[t]
\hskip 1cm \psfig{figure=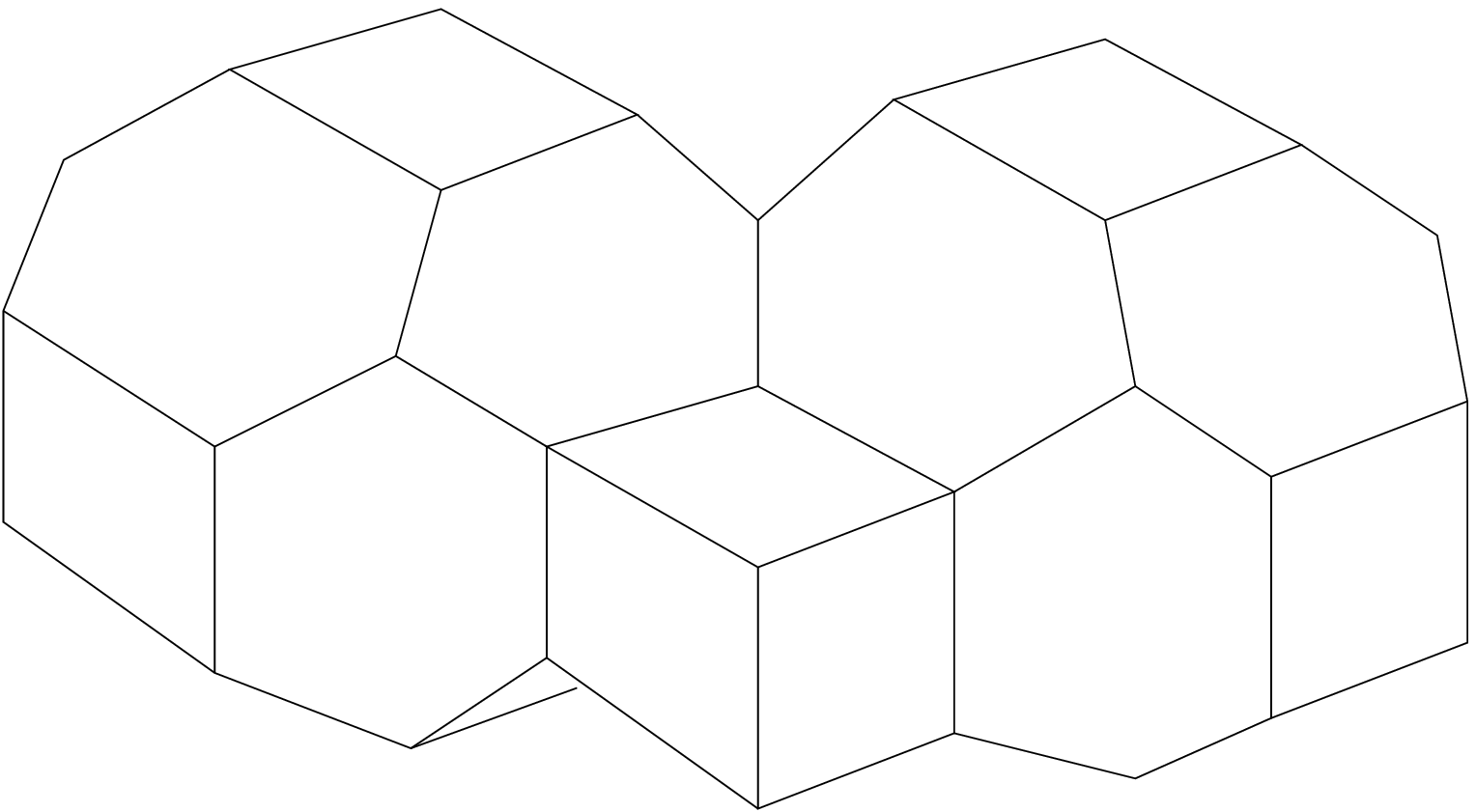,height=6cm}
\caption{}
\label{fig7}
\end{figure}
\end{center}

All the external 36 plaquettes of this diagram are labeled with
$j=\frac{1}{2}$ while the 3 internal (shared) plaquettes
are labeled with $j=1$; then it gives a power of $t^{
\frac{3}{2} \times 36 + 4 \times 3}= t^{66}$.  
We have computed the strong coupling expansion of $f$
up to power 53 in $t$ which involves
34 diagrams grouped in 18 different powers of $t$: 

\begin{eqnarray}
f=4t^9-8t^{18}+9t^{24}+76/3t^{27}-12t^{33}-160t^{36}+
72t^{37}+60t^{39} \nonumber \\
 -432t^{42}+360t^{43}+8224/5t^{45}+612t^{46}-
 1728t^{47}-2961/2t^{48} \nonumber \\
    +720t^{49}+5052t^{51}-8640t^{52}+2664t^{53}
%-17480/3t^{54}
\label{eq:f}
\end{eqnarray}

We plot in FIG. 8 the $f$ vs. $\beta$ for different powers
of truncation of the expansion.   

\begin{center}
\begin{figure}[t]
\hskip 1cm \psfig{figure=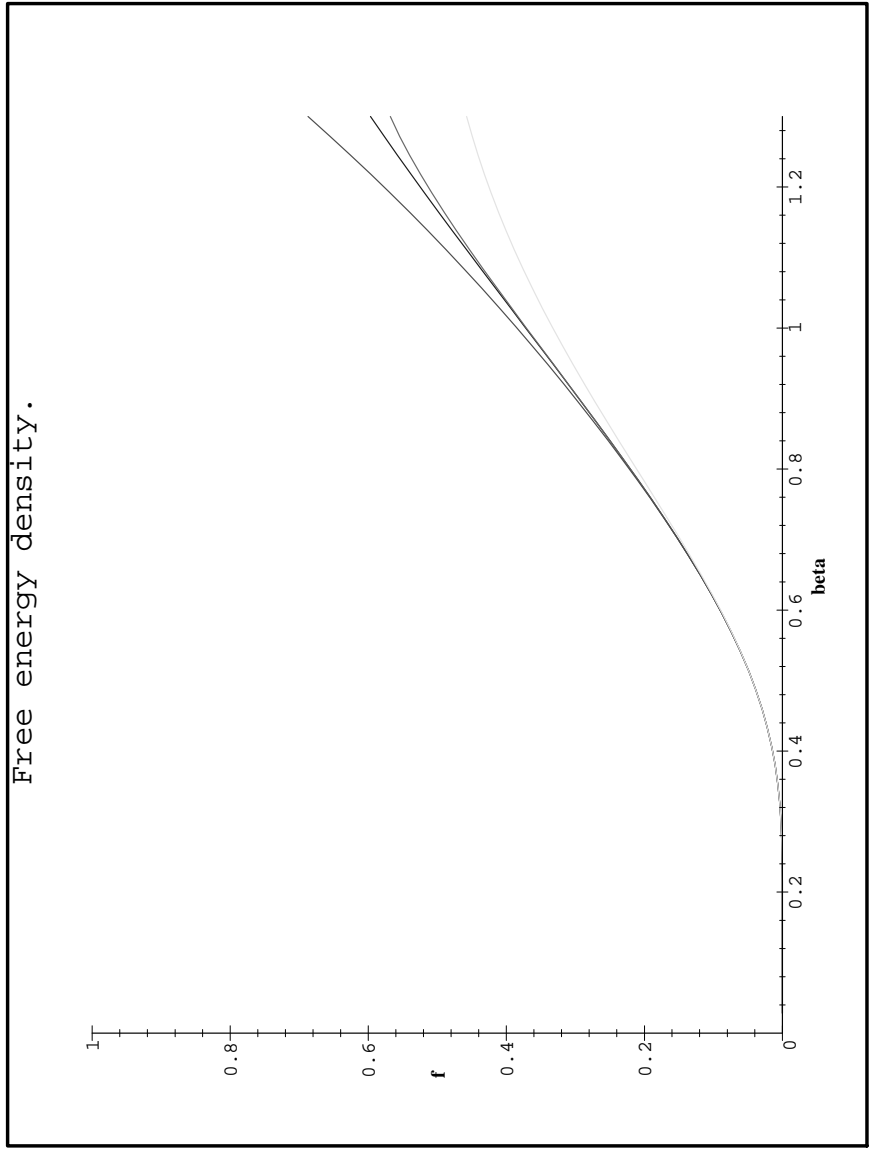,width=13cm}
\caption{Free energy $f$ vs. $\beta$ for several truncation 
orders. From above:
$t^{27}$, $t^{39}$, $t^{46}$, $t^{53}$ and $t^{18}$.}
\label{fig8}
\end{figure}
\end{center}

As long as we enter in the weak coupling regime 
($\beta > 1$), 
one can appreciate a clear difference with the series expansions 
of ref. \cite{dz83} which were performed in an ordinary 
cubic lattice and with a different
truncation criteria (they consider diagrams 
up to 16 plaquettes which corresponds to $t^{24}$). 
The explanation of this difference relies on the fact 
that in the strong coupling expansion in a cubic lattice up
to 16 plaquettes all the terms except two are positive
while in our case we have almost equal number of contributions    
with both signs.

Thus, one can observe that the introduction of the dual of the
tethraedral lattice, besides simplifiying the strong
coupling computations provides a straightforward procedure
to obtain the desired terms of the series expansion.
This turns to be an advantage in order to reach the
weak coupling regime.

\section{CONCLUSIONS}

We have introduced a Hamiltonian spin network representation for a SU(2)
lattice gauge theory. This gauge invariant representation is given in
terms of an independent basis that diagonalize the electric part of the
Hamiltonian. The corresponding Lagrangian formulation is also developed.
This formulation takes a purely geometrical form in terms of sums over
coloured surfaces and allow to combine the powerfull Lagrangian
techniques with the redundancy free description typical of the loop
representation. This action may be written on a thetrahedral lattice 
explicitly in
terms of the Racah coefficients. 
The computation of group theoretical factors of diagrams of the 
strong coupling expansion of high orders of $t$ becomes
straigtforward, only involving these Racah coefficients.
Also, we have a compact expression for the coloured surfaces 
action which allows to perform numerical computations.

In the naive weak coupling limit, the
area dependent factors become equal to 1 and the action is 
purely topological.
One can immediately check that this limit correspond to the Ouguri
\cite{O} form of the B-F topological field theory, which in three
dimensions is known \cite{W} to coincide with the pure gravity action.
In this case the use of the Biedenharn and Elliot identity
\cite{be}  allows to show that the 
action is invariant under the renormalization group. Thus, in the 
different context of QCD,
this suggests that 
the Yang Mills action in terms of coloured surfaces may be particularly 
well suited for the study of the effective theories.

Even though the method developed here was for SU(2) in 2+1 dimensions,
the extension to other groups, in particular to SU(3), is
straightforward. The corresponding spin networks would simply carry the
quantum numbers required to charactherize the irreducible
representations of the Lie group under study. It is also possible to
extend this formulation to the four dimensional case, by making use of
the higher order Racah-Wigner $j$-coefficients. An important simplification of
the path integral (\ref{eq:spine2}) with the same weak coupling regime could be
obtained by making use of the Ponzano and Regge asymptotic form of the
Racah-Wigner $j$-symbols. We hope to present elsewhere a more detailed
analysis of these developments.

\end{document}